# The Limits of Fair Medical Imaging AI In The Wild

Yuzhe Yang[1]*†, Haoran Zhang[1]*, Judy W. Gichoya[2],
Dina Katabi[1], Marzyeh Ghassemi[1,3]

**Affiliations:**

[1] Department of Electrical Engineering and Computer Science, Massachusetts Institute of Technology; Cambridge, MA, USA.
[2] Department of Radiology, Emory University School of Medicine; Atlanta, GA, USA.
[3] Institute for Medical Engineering and Science, Massachusetts Institute of Technology; Cambridge, MA, USA.

\* Co-first authors
† Corresponding author. Email: yuzhe@mit.edu

## Abstract

As artificial intelligence (AI) rapidly approaches human-level performance in medical imaging, it is crucial that it does not exacerbate or propagate healthcare disparities. Prior research has established AI's capacity to infer demographic data from chest X-rays, leading to a key concern: do models using demographic shortcuts have unfair predictions across subpopulations? In this study, we conduct a thorough investigation into the extent to which medical AI utilizes demographic encodings, focusing on potential fairness discrepancies within both in-distribution training sets and external test sets. Our analysis covers three key medical imaging disciplines: radiology, dermatology, and ophthalmology, and incorporates data from six global chest X-ray datasets. We confirm that medical imaging AI leverages demographic shortcuts in disease classification. While correcting shortcuts algorithmically effectively addresses fairness gaps to create "locally optimal" models within the original data distribution, this optimality is not true in new test settings. Surprisingly, we find that models with less encoding of demographic attributes are often most "globally optimal", exhibiting better fairness during model evaluation in new test environments. Our work establishes best practices for medical imaging models which maintain their performance and fairness in deployments beyond their initial training contexts, underscoring critical considerations for AI clinical deployments across populations and sites.

# Main

As AI models are increasingly deployed in real-world clinical settings[1,2], it is crucial to evaluate not only model performance, but also potential biases towards specific demographic groups[3,4]. While deep learning has achieved human-level performance in numerous medical imaging tasks[5-7], existing literature indicates a tendency for these models to manifest existing biases in the data, causing performance disparities between protected subgroups[8-13]. For instance, chest X-ray classifiers trained to predict the presence of disease systematically underdiagnose Black patients[14], potentially leading to delays in care. To ensure the responsible and equitable deployment of such models, it is essential to understand the source of such biases and, where feasible, take actions to correct them[15,16].

Recent studies have unveiled the surprising ability of deep models to predict demographic information such as self-reported race[17], sex, and age[18] from medical images, achieving performance far beyond that of radiologists. These insights raise the concern of disease prediction models leveraging demographic features as heuristic "shortcuts" [19,20] – correlations that are present in the data, but have no real clinical basis[20], for instance deep models using the hospital as a shortcut for disease prediction[21,22].

In this work, we investigate four questions. First, whether disease classification models also utilize *demographic information* as shortcuts, and whether such demographic shortcuts result in biased predictions. Second, we evaluate the extent to which state-of-the-art methods can remove such shortcuts and create "locally optimal" models that are also fair. Third, we consider real-world clinical deployments settings where shortcuts may not be valid in the out-of-distribution data, in order to dissect the interplay between algorithmic fairness and shortcuts when data shifts. Finally, we explore which algorithms and model selection criteria can lead to "globally optimal" models that maintain fairness when deployed in an out-of-distribution setting.

We perform a systematic investigation into how medical AI leverages demographic shortcuts through these questions, with an emphasis on fairness disparities across both in-distribution training and external test sets. Our primary focus is on chest X-ray (CXR) prediction models, with further validation in dermatology (Extended Data Fig. 1) and ophthalmology (Extended Data Fig. 2). Our X-ray analysis draws upon six extensive, international radiology datasets: MIMIC-CXR[23], CheXpert[24], NIH[25], SIIM[26], PadChest[27], and VinDr[28]. We explore fairness within both individual and intersectional subgroups spanning race, sex, and age[14]. Our assessment uncovers compelling new insights on how medical AI encodes demographics, and the impact this has on various fairness considerations, especially when models are applied outside their training context during real-world domain shifts, with actionable insights on what models to select for fairness under distribution shift.

**Table 1. Demographic and label characteristics of the six X-ray datasets used in this study.**

|  |  | MIMIC | CheXpert | NIH | SIIM | PadChest | VinDr |
|---|---|---|---|---|---|---|---|
|  | Location | Boston, MA | Stanford, CA | Bethesda, MD | Bethesda, MD | Alicante, Spain | Hanoi, Vietnam |
|  | # Images | 357,167 | 222,792 | 112,120 | 11,582 | 144,478 | 6,354 |
|  | % Frontal | 64.5 | 85.5 | 100.0 | 100.0 | 69.1 | 100.0 |
|  | Sample Image | 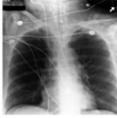 | 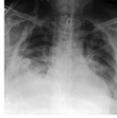 | 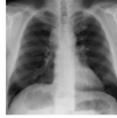 | 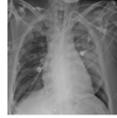 | 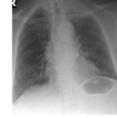 | 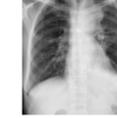 |
| Sex (%) | Female | 47.8 | 40.7 | 43.5 | 44.6 | 50.4 | 43.1 |
|  | Male | 52.2 | 59.3 | 56.5 | 55.4 | 49.6 | 56.9 |
| Race (%) | Asian | 3.1 | 10.5 | - | - | - | - |
|  | Black | 15.6 | 5.4 | - | - | - | - |
|  | White | 61.0 | 56.4 | - | - | - | - |
|  | Other | 20.3 | 27.8 | - | - | - | - |
| Age (%) | 0-18 | - | - | 4.8 | 5.0 | 3.7 | 21.8 |
|  | 18-40 | 13.8 | 13.9 | 27.7 | 27.3 | 9.2 | 16.0 |
|  | 40-60 | 31.1 | 31.1 | 43.9 | 42.9 | 26.5 | 27.1 |
|  | 60-80 | 40.0 | 39.0 | 22.7 | 23.9 | 38.0 | 30.0 |
|  | 80-100 | 15.1 | 16.0 | 0.9 | 0.9 | 22.6 | 5.1 |
| Intersection (%) | Asian Female | 1.5 | 4.5 | - | - | - | - |
|  | Asian Male | 1.6 | 6.0 | - | - | - | - |
|  | Black Female | 9.3 | 2.6 | - | - | - | - |
|  | Black Male | 6.3 | 2.7 | - | - | - | - |
|  | White Female | 27.3 | 22.2 | - | - | - | - |
|  | White Male | 33.8 | 34.1 | - | - | - | - |
|  | Others Female | 9.8 | 11.3 | - | - | - | - |
|  | Others Male | 10.5 | 16.5 | - | - | - | - |
| Task Prevalence (%) | No Finding | 39.8 | 10.0 | 53.8 | - | 34.9 | 41.2 |
|  | Effusion | 20.0 | 38.6 | 11.9 | - | 5.9 | 7.5 |
|  | Pneumothorax | 3.4 | 8.7 | 4.7 | 28.4 | 0.3 | 0.7 |
|  | Cardiomegaly | 14.9 | 12.1 | 2.5 | - | 9.5 | 22.6 |

## Results

**Datasets and Model Training**

We utilize six publicly available chest X-ray datasets as described in Table 1. We focus on four binary classification tasks that have been shown to have disparate performance between protected groups[8,29]: "No Finding", "Effusion", "Pneumothorax", and "Cardiomegaly". The detailed prevalence rates of the diseases for each demographic subgroup are in Extended Data Table 1.

We also examine medical AI applications in dermatology and ophthalmology. Specifically, we use the ISIC dataset[30] with "No Finding" as the task for dermatological imaging (Extended Data Fig. 1a), and the ODIR dataset[31] with "Retinopathy" as the task for ophthalmology images (Extended Data Fig. 2a).

To evaluate fairness, we examine the class-conditioned error rate that is likely to lead to worse patient outcomes for a screening model. For No Finding, a false positive indicates falsely predicting a patient is healthy when they are ill, which could lead to delays in treatment[14]; we therefore evaluate the differences in False Positive Rate (FPR) between demographic groups. For all other diseases, we evaluate the False Negative Rate (FNR) for the same reason. Equality in these metrics is equivalent to equality of opportunity[32].

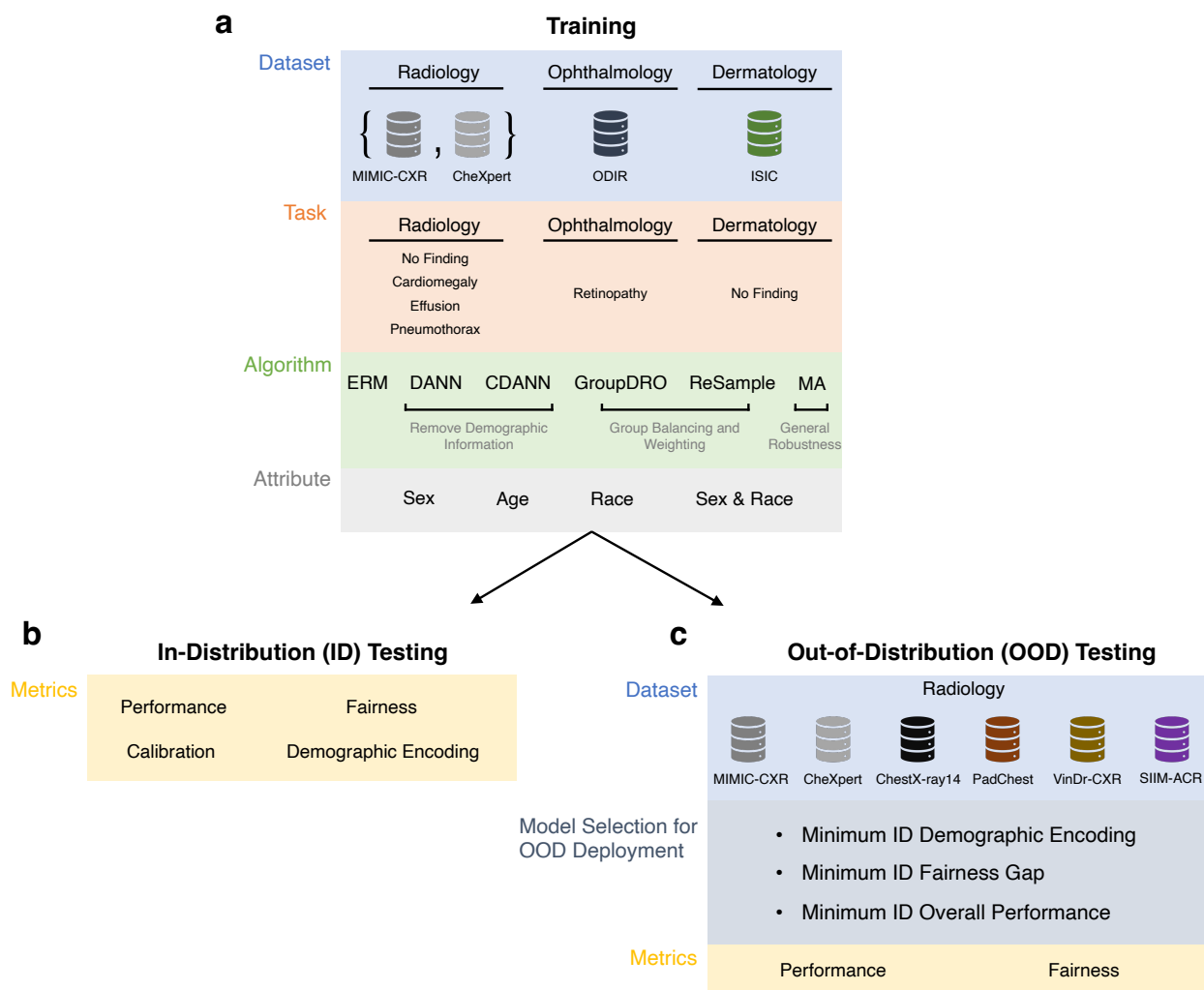

**Figure 1. Overall experimental pipeline. a,** We train a grid of deep learning models on medical images from a variety of modalities on several clinical tasks. We apply a variety of state-of-the-art algorithms to mitigate shortcuts, for up to four demographic attributes (where available). **b,** We evaluate each model in-distribution (i.e., on the same dataset where it is trained), along the axis of performance, fairness, amount of demographic encoded, and calibration. **c,** We evaluate the performance and fairness of chest X-ray classification models on out-of-distribution (OOD) domains. To mimic a realistic deployment setting where OOD samples are not observed, we choose the "best" model based on several in-distribution selection criteria.

We train a grid of deep convolutional neural networks[33] on MIMIC-CXR (radiology), ODIR (ophthalmology), and ISIC (dermatology), varying the classification task. Our approach follows prior work which achieves state-of-the-art performance in these tasks[9,14] using standard training or Empirical Risk Minimization (ERM)[34]. We also evaluate five algorithms designed to remove spurious correlations, or increase model fairness during training. We categorize these algorithms into those that (1) reweight samples based on their group to combat underrepresentation (ReSample[35], GroupDRO[36]), (2) adversarially remove group information from model representations (DANN[37], CDANN[38]), and (3) more generically attempt to improve model generalization (MA[39]). In total, our analysis encompassed a total of 3,456 models trained on MIMIC-CXR, corresponding to the cartesian product of 4 tasks, 4 demographic attributes, 6 algorithms, 12 hyperparameter settings, and 3 random seeds.

**Algorithmic encoding of protected attributes leads to model fairness gaps**

We separately train deep learning models for our four distinct CXR prediction tasks ("No Finding", "Cardiomegaly", "Effusion", "Pneumothorax"), as well as "Retinopathy" in ophthalmology, and "No Finding" in dermatology. Each model consists of a feature extractor followed by a disease prediction head. We then employ a transfer learning approach, wherein we keep the weights of the feature extractor frozen and retrain the model to predict sensitive attributes (e.g., race). This allows us to assess the amount of attribute-related information present in the features learned by each model as measured by the area under the ROC curve (AUC) for attribute prediction (details in the Methods section). We extend prior work[40] demonstrating that deep models trained for disease classification encode demographic attributes, and test across a wider range of settings. As Figs. 2a, 2c, and 2e confirms, the penultimate layer of different disease models contains significant information about four demographic attributes (age, race, sex, and the intersection of sex and race), and that is consistent across different tasks and medical imaging modalities.

We then assess the fairness of these models across demographic subgroups as defined by equal opportunity[32], i.e., discrepancies in the model's false negative rate (FNR) or false positive rate (FPR) for demographic attributes. We focus on underdiagnosis[14], i.e., discrepancies in FPR for "No Finding" and discrepancies in FNR for other diseases. For each demographic attribute, we identify two key subgroups with sufficient sample sizes: age groups "80-100" (n=8,063) and "18-40" (n=7,319); race groups "White" (n=32,732) and "Black" (n=8,279); sex groups "female" (n=25,782) and "male" (n=27,794); sex & race groups "White male" (n=18,032) and "Black female" (n=5,027). In all tasks, we observe that the models displayed biased performance within the four demographic attributes, as evidenced by the FNR disparities (Fig. 2b). The observed gaps can be as large as 30% for age. The same results hold for the other two imaging modalities (Figs. 2d, 2f). Similar results for overdiagnosis (FNR of No Finding, and FPR for disease prediction) can be found in Extended Data Fig. 3.

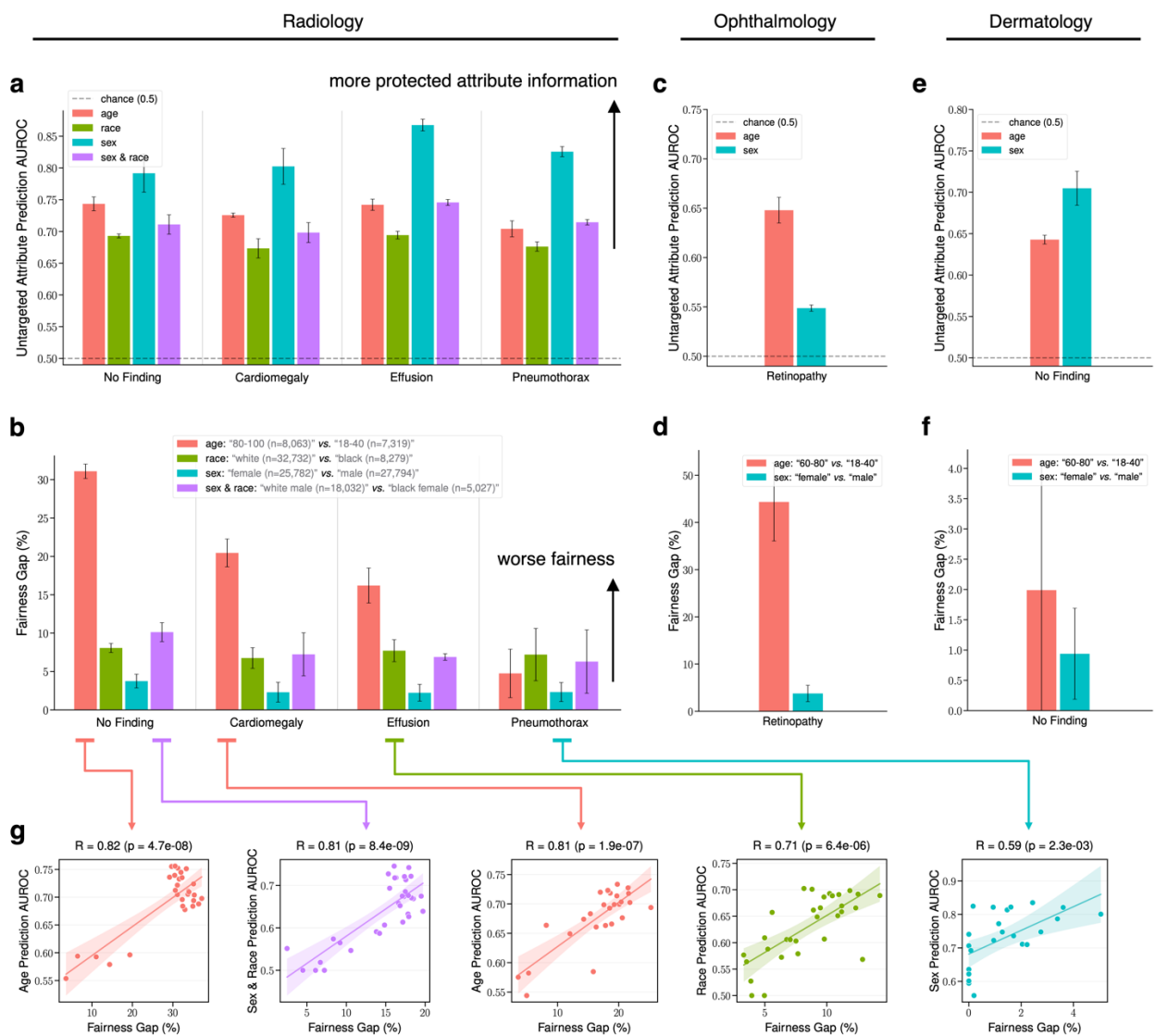

**Figure 2. Medical imaging models encode sensitive attributes and are unfair across subgroups. a,** The area under the ROC curve (AUROC) of demographic attribute prediction from frozen representations for the best ERM model. We train ERM models on MIMIC-CXR to predict four different binary tasks. ERM representations encode demographic attributes to a high degree. **b,** The fairness gap, as defined by the FPR gap for No Finding, and the FNR gap for all other tasks for the best ERM model. ERM models exhibit high fairness gaps, especially between age groups. **c,** The AUROC of demographic attribute prediction from frozen representations for the best ERM model on the ODIR dataset (ophthalmology), following the same experimental setup. **d,** The fairness gap for the best ERM model on the ODIR dataset (ophthalmology). **e,** The AUROC of demographic attribute prediction from frozen representations for the best ERM model on the ISIC dataset (dermatology), following the same experimental setup. **f,** The fairness gap for the best ERM model on the ISIC dataset (dermatology). **g,** The correlation between attribute prediction performance and fairness for all learned models. We exclude models with suboptimal performance, i.e., with an overall validation AUROC below 0.7. The attribute prediction AUROC shows a high correlation with the fairness gap (No Finding, age: R=0.82, p=4.7e-08; No Finding, sex & race: R=0.81, p=8.4e-09; Cardiomegaly, age: R=0.81, p=1.9e-07; Effusion, race: R=0.71, p=6.4e-06; Pneumothorax, sex: R=0.59, p=2.3e-03).

We further investigate the degree to which demographic attribute encoding "shortcuts" may impact model fairness. We note that a model encoding demographic information does not necessarily imply a fairness violation, as the model may not necessarily use this information for its prediction. For each task and attribute combination, we train different models with varying hyperparameters (see Methods). We focus on the correlation between the degree of encoding of different attributes, and the fairness gaps as assessed by underdiagnosis. Fig. 2g shows that a stronger encoding of demographic information is significantly correlated with stronger model unfairness (No Finding, age: R=0.82, p=4.7e-08; No Finding, sex & race: R=0.81, p=8.4e-09; Cardiomegaly, age: R=0.81, p=1.9e-07; Effusion, race: R=0.71, p=6.4e-06; Pneumothorax, sex: R=0.59, p=2.3e-03). Such consistent observations indicate that models using demographic encodings as heuristic shortcuts also have larger performance disparities.

**Mitigating shortcuts creates locally optimal models that are fair and performant**

We perform model evaluations first in the in-distribution (ID) setting, where ERM models trained and tested on data from the same source perform well. We compare ERM to state-of-the-art robustness methods that have been designed to effectively address fairness gaps while maintaining overall performance. As shown in Fig. 3a, ERM models exhibit large fairness gaps across age groups when predicting Cardiomegaly (i.e., models centered in the top right corner, FNR gap 20% between groups "80-100" and "18-40"). By applying debiasing robustness methods that correct demographic shortcuts, such as GroupDRO and DANN, the resulting models are able to close the FNR gap, while achieving similar AUROCs (e.g., the bottom right corner). Our results hold when using the worst group AUROC as the performance metric (Extended Data Fig. 4), and across different combinations of diseases and attributes (Figs. 3b, Extended Data Fig. 4).

To demonstrate the value of model debiasing, we further plot the set of *locally optimal models* - those on the Pareto front[41] that balance the performance-fairness tradeoff most optimally on ID data (Fig. 3a). Those models that lie on this front are "locally optimal", as they have the smallest fairness gap that can be achieved for a fixed performance constraint (e.g., AUROC > 0.8). In the ID setting, we find several existing algorithms that consistently achieve high ID fairness without losing overall performance for disease prediction (Figs. 3a, 3b, Extended Data Fig. 4).

Similar to our observations in radiology, we identify fairness gaps within subgroups based on age and sex in dermatology and ophthalmology, respectively (Figs. 2d, 2f). We further verify the Pareto front for both attributes, where similar observations hold that algorithms for fixing demographic shortcuts could improve in-distribution fairness while incurring minimal detriments to performance (Figs. 3c, 3d).

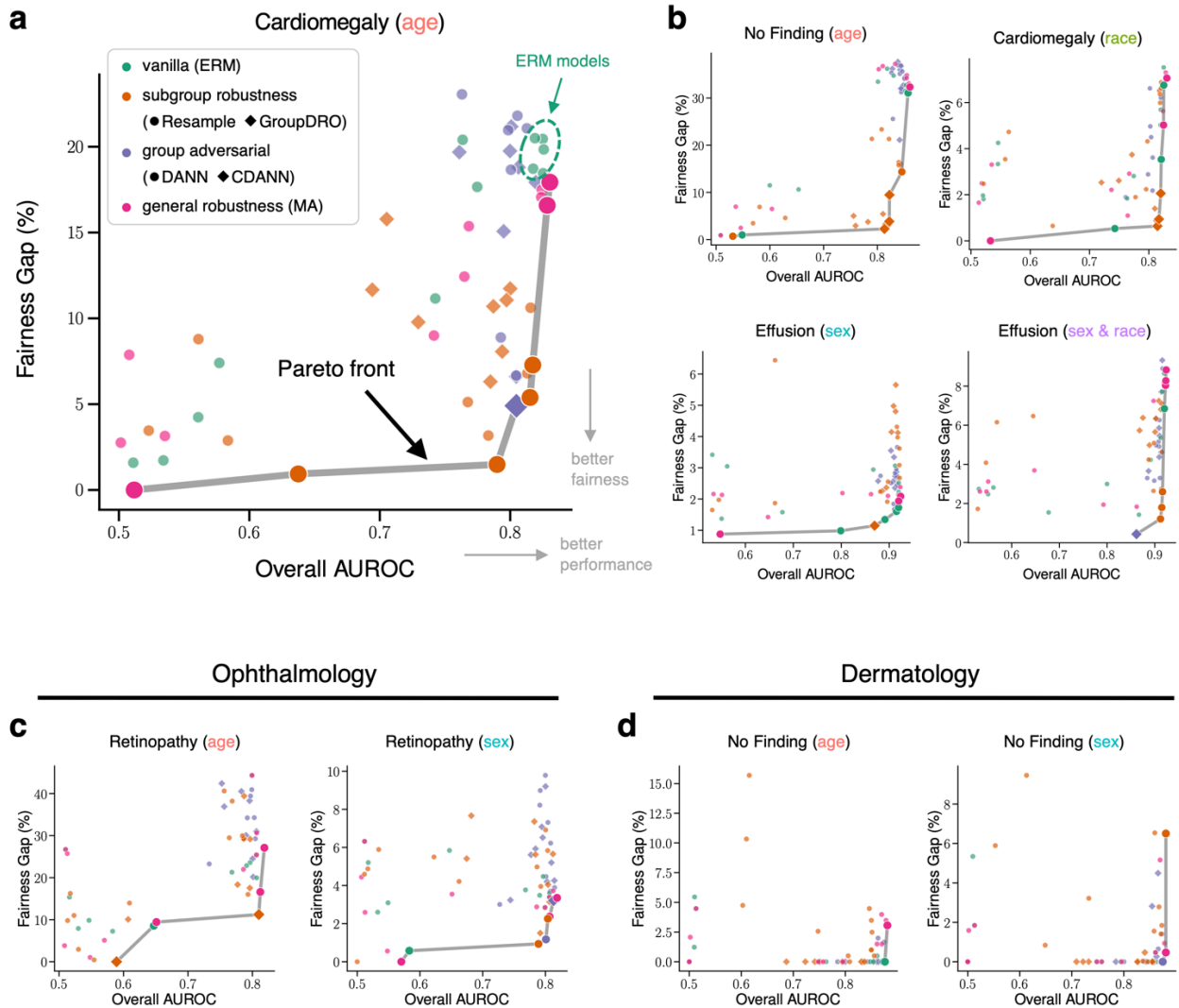

**Figure 3. Algorithms for removing demographic shortcuts mitigate in-distribution fairness gaps and maintain performance. a, b,** Trade-off between the fairness gap and overall AUROC for all trained models. Each plot represents a specific disease prediction task (e.g., Cardiomegaly) with a specific attribute (e.g., age). In each case, we plot the Pareto front, the best achievable fairness gap with a minimum constraint on the performance. **c, d,** Trade-off between the fairness gap and the overall AUROC on the ODIR dataset (ophthalmology) and ISIC dataset (dermatology).

**Locally optimal models exhibit trade-offs in other metrics**

We examine how locally optimal models that balance fairness and AUROC impact other metrics, as previous work has shown it is theoretical impossibility to balance fairness measured by probabilistic equalized odds and calibration by group[42,43]. We find that optimizing fairness alone leads to worse results for other clinically meaningful metrics in some cases, indicating an inherent tradeoff between fairness and other metrics. First, for the "No Finding" prediction task, enforcing fair predictions across groups results in worse expected calibration error gap (ECE Gap, Fig. 4) between groups. Across different demographic attributes, we find a consistent statistically significant negative correlation between ECE Gap and Fairness Gap (age: R=-0.85, p=7.5e-42; race: R=-0.64, p=6.1e-15; sex: R=-0.73, p=4.4e-28; sex & race: R=-0.45, p=1.9e-08).

We explore the relationship between fairness and other metrics, including average precision and average F1 score. For "No Finding" prediction, fairer models lead to both worse average precision and F1 score (Extended Data Fig. 5a). The same trend holds across different diseases, e.g., for Effusion (Extended Data Fig. 5b). These findings stress that these models, though being locally optimal, exhibit worse results on other important and clinically relevant performance metrics. This uncovers the limitation of blindly optimizing fairness, emphasizing the necessity for more comprehensive evaluations to ensure the reliability of medical AI models.

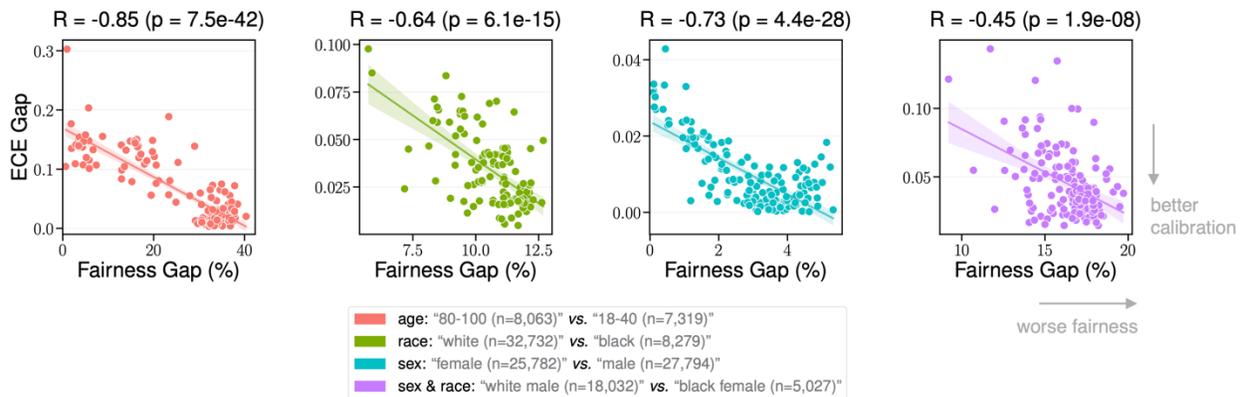

**Figure 4. The tradeoff between the fairness gap and the expected calibration error (ECE) gap.** For the No Finding task, we examine the trade-off between the fairness gap and the ECE gap (age: R=-0.85, p=7.5e-42; race: R=-0.64, p=6.1e-15; sex: R=-0.73, p=4.4e-28; sex & race: R=-0.45, p=1.9e-08). Correlations for additional metrics can be found in Extended Data Fig. 5.

**Locally optimal model fairness does not transfer under distribution shift**

When deploying AI models in real settings, it is crucial to ensure that models can generalize to data from unseen institutions or environments. We directly test all trained models in the out-of-distribution (OOD) setting, where we report results on external test datasets that are unseen during model development. Fig. 5 illustrates that the correlation between ID and OOD performance is high across different settings, which has been observed in prior work[44,45]. However, we find that there is no consistent correlation between ID and OOD fairness. For example, Fig. 5b shows an instance where the correlation between ID fairness and OOD fairness is strongly positive ("Effusion" with "age" as the attribute; R=0.98, p=3.0e-36), while Fig. 5c shows an instance where the correlation between these metrics is actually significantly negative ("Pneumothorax" with "sex & race" as the attribute; R=-0.50, p=4.4e-03). Across 16 combinations of task and attribute, we find that 5 such settings exhibit this negative correlation, and 3 additional settings exhibit only a weak (R < 0.5) positive correlation (see Extended Data Figs. 6a, 6b for additional correlation plots). Thus, improving ID fairness may not lead to improvements in OOD fairness, highlighting the complex interplay between fairness and distribution shift[46,47].

In addition, we investigate whether models achieving ID Pareto optimality between fairness and performance will maintain in OOD settings. As shown for "Cardiomegaly" prediction using race as the attribute, models originally on the Pareto front ID (Fig. 5d) do not guarantee to maintain Pareto optimality when deployed in a different OOD setting (Fig. 5e). We show additional examples of this phenomenon in Extended Data Fig. 6c.

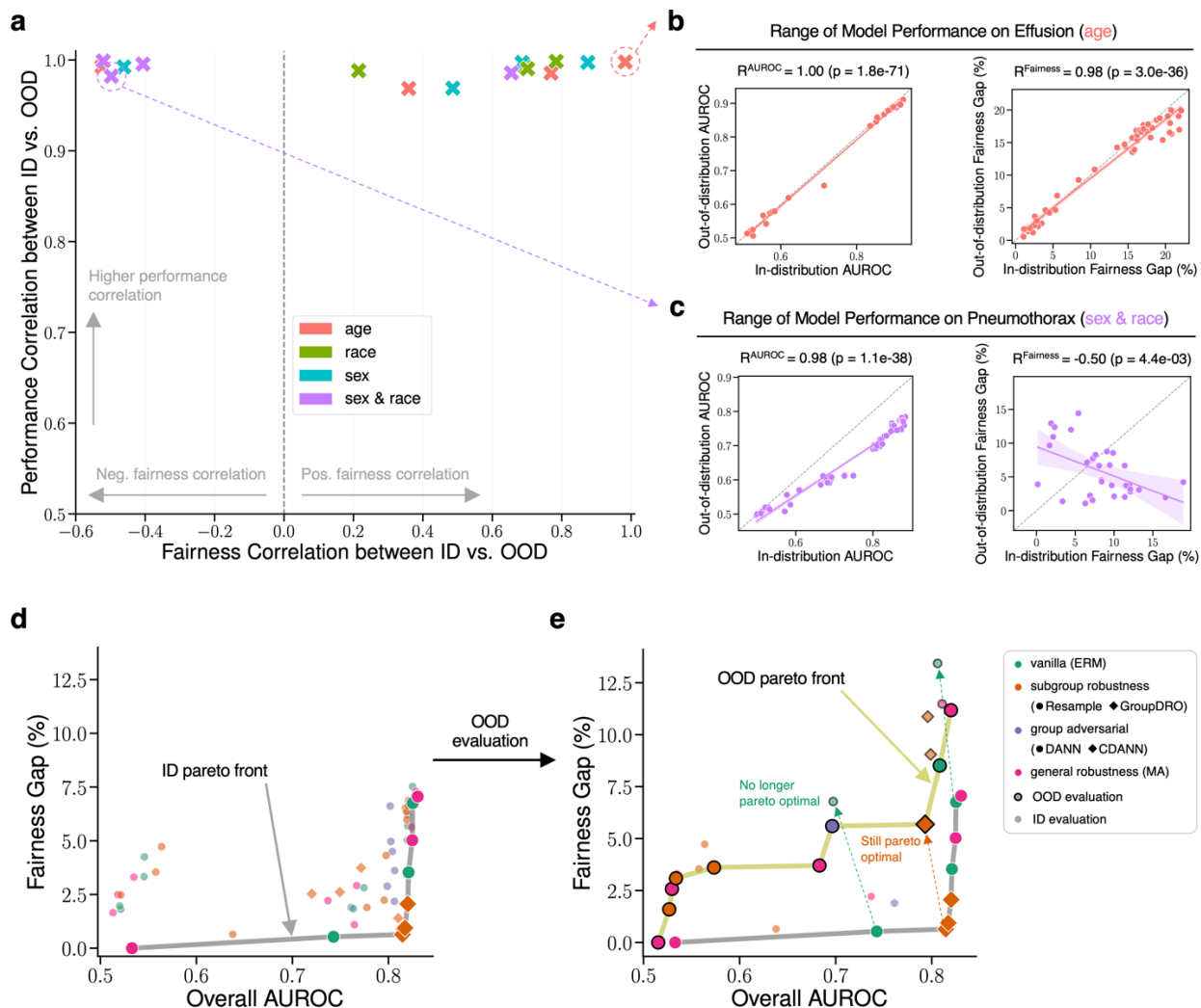

**Figure 5. The transfer of performance (overall AUROC) and fairness between the ID (MIMIC-CXR) and OOD datasets. a,** We plot the Pearson correlation coefficient of ID vs. OOD performance versus the Pearson correlation coefficient of ID vs. OOD fairness. Here, each point is derived from a grid of models trained on a particular combination of task and attribute. We find that there is a high correlation between ID and OOD performance in all cases, but the correlation between ID and OOD fairness is tenuous. **b, c,** We show how two particular points in the first plot are obtained; one where fairness transfers ("Effusion" with "age" as the attribute; R=0.98, p=3.0e-36), and one where it does not ("Pneumothorax" with "sex & race" as the attribute; R=-0.50, p=4.4e-3). **d, e,** We show the transformation of the ID Pareto front to the OOD Pareto front, for cardiomegaly prediction using race as the attribute, finding that models that are Pareto optimal ID often do not maintain Pareto optimality OOD.

**Dissecting model fairness under distribution shift**

To disentangle the OOD fairness gap, we present a way to decompose model fairness under distribution shift. Specifically, we decompose and attribute the change in fairness between ID and OOD to be the difference in performance change for each of the groups, i.e., the change in fairness is determined by how differently the distribution shift affects each group (details in Methods).

In Fig. 6, we show an example of transferring a model to predict No Finding trained on CheXpert (ID) to the MIMIC-CXR data (OOD), while evaluating fairness across genders. We find that the model is fair with respect to the FPR gap in the ID setting (-0.1% gap, not significant), but has a significant FPR gap when deployed in the OOD setting (3.2%), with females being underdiagnosed at a higher rate (Fig. 6b).

We then segment this FPR gap by gender, and find that females experience an increase in FPR of 3.9%, while males experience an increase in FPR of 0.8% (Fig. 6c). In other words - the model becomes worse for both groups in an OOD setting, but to a much larger extent for female patients. This decomposition suggests that mitigation strategies which reduce the impact of the distribution shift on females could be effective in reducing the OOD fairness gap in this instance. We further extend this study to a larger set of tasks and protected attributes (Extended Data Fig. 7). Across all settings, the disparate impact of distribution shift on each group is a significant component, indicating that mitigating the impact of distribution shift is as important as mitigating ID fairness, if the goal is to achieve a fair model out-of-distribution.

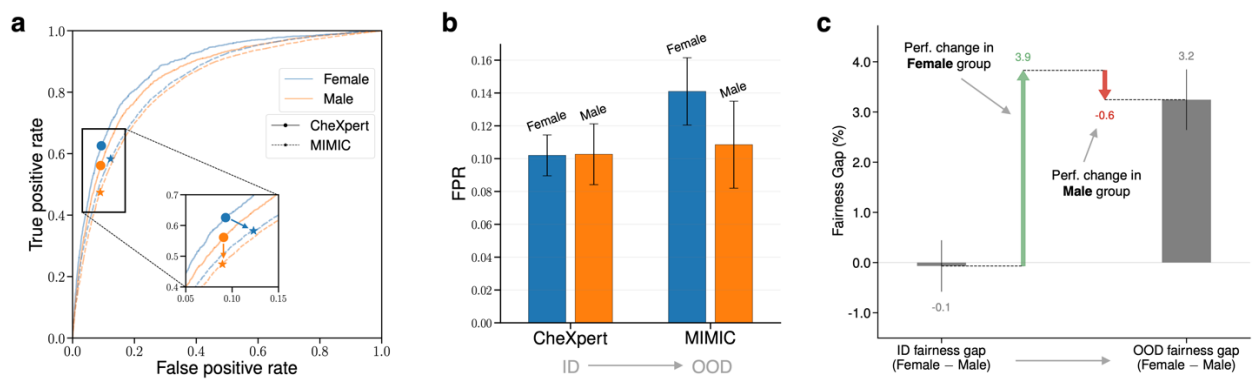

**Figure 6. Examining the gender biases of an ERM model for No Finding prediction, trained on CheXpert and deployed on MIMIC. a,** We plot the receiver operating characteristic (ROC) curves for each group and each dataset, marking the operating point of the model. **b,** This shift in the operating point results in a change in the FPR values of males and females on the OOD dataset, with both groups seeing increased underdiagnosis, but females are impacted more heavily. **c,** We decompose the OOD fairness gap as a function of the ID fairness gap, and the change in FPR for each of the groups, finding that the large increase in OOD fairness is primarily attributable to the increase in FPR for females.

**Globally optimal model selection for out-of-domain fairness**

Fig. 5 shows that selecting a model based on in-distribution fairness may not lead to a model with optimal OOD fairness. Here, we examine alternate model selection criteria that may lead to better OOD fairness, when we only have access to ID data. Our goal is to find "***globally optimal***" models which maintain their performance and fairness in new domains. First, we subset our selection only to models that have satisfactory ID overall performance (defined as those with overall validation AUROC no less than 5% of the best ERM model). As validated in Fig. 5, this set of models will also have satisfactory OOD performance.

Next, we propose eight candidate model selection criteria (Fig. 7a), corresponding to selecting the model from this set that minimizes or maximizes some in-distribution metric. We evaluate the selected model by its OOD fairness across five external datasets, each containing up to four attributes and up to four tasks, corresponding to a total of 42 settings. We compare the OOD fairness of the selected model to the OOD fairness of an "oracle", which observes samples from the OOD dataset and directly chooses the model with the smallest OOD fairness gap. For each setting, we compute the increase in fairness gap of each selection criteria relative to the oracle. In Fig. 7a, we report the mean across the 42 settings, as well as the 95% confidence interval computed from 1,000 bootstrap iterations. We find that, surprisingly, selecting the model with the minimum ID fairness gap may not be optimal. Instead, two other criteria based on selecting models where the embedding contains the least attribute information, lead to a lower average OOD fairness gap. For instance, we observe a significantly lower increase in OOD fairness gap by selecting models with the "Minimum Attribute Prediction Accuracy" as compared to "Minimum Fairness Gap" ($p=9.60e-94$, one-tailed Wilcoxon rank-sum test). The result echoes our finding in Fig. 2 that the encoding of demographic attributes is positively correlated with ID fairness.

Finally, we study the fairness of each algorithm in the OOD setting. We maintain the performance cutoff described above, and select the model for each algorithm with the lowest ID fairness gap. In Fig. 7b, we report the mean increase in OOD fairness gap relative to the oracle across the same 42 settings. We find that methods which remove demographic information from embeddings (specifically, DANN) lead to the lowest average OOD fairness gap ("DANN" vs "ERM": $p=1.86e-117$, one-tailed Wilcoxon rank-sum test). Our findings demonstrate that evaluating and removing demographic information encoded by the model in-distribution may be the key to "globally optimal" models which transfer both performance and fairness to external domains.

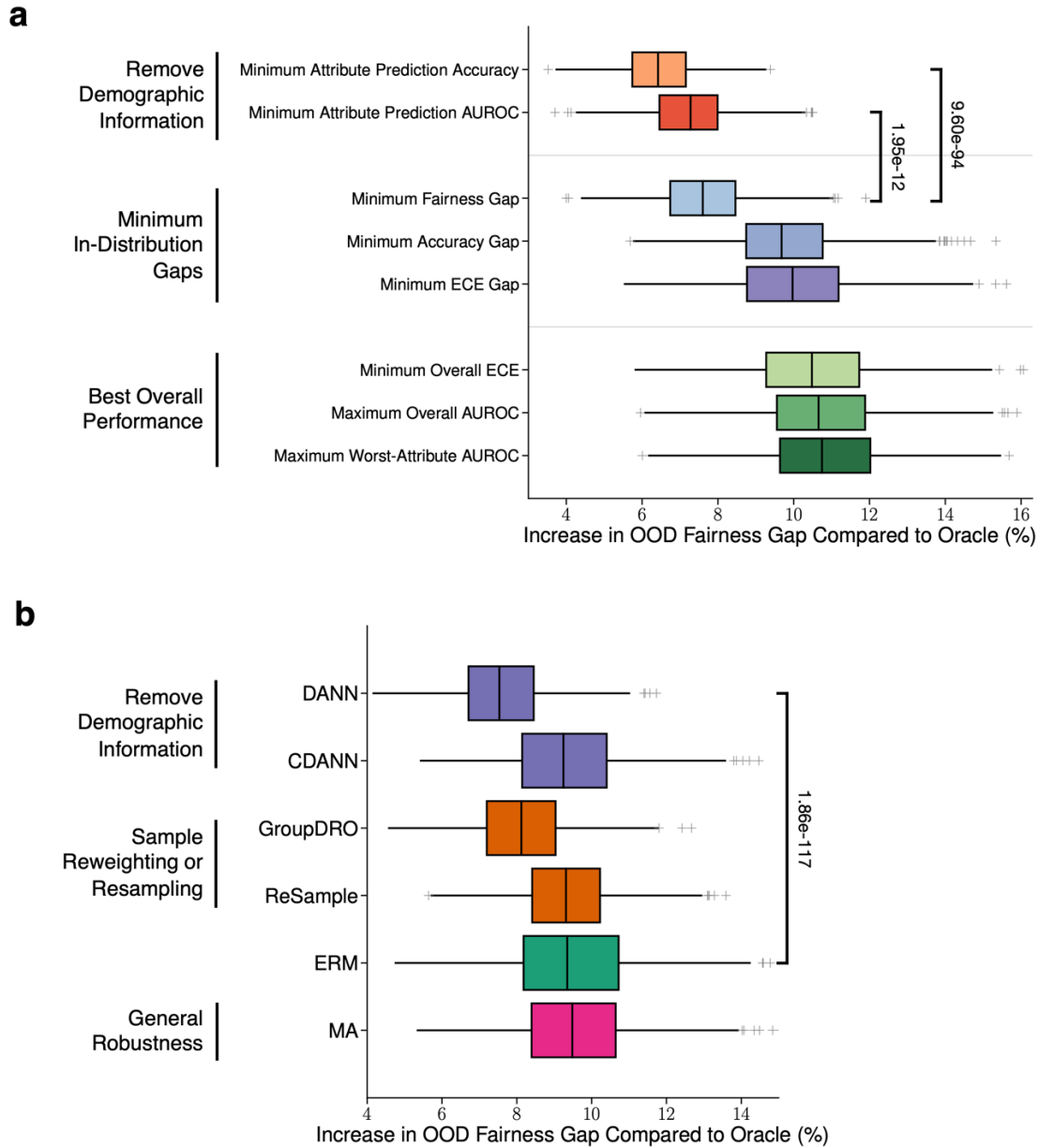

**Figure 7. OOD fairness of models with different model selection criteria and for different algorithms. a,** We vary the in-distribution model selection criteria, and compare the selected model against the oracle which chooses the model that is most fair OOD. We plot the increase in OOD fairness gap of the selected model over the oracle, averaged across 42 combinations of OOD dataset, task, and attribute. We find that selection criteria based on choosing models with minimum attribute encoding achieve better OOD fairness than naively selecting based on in-distribution fairness, or other aggregate performance metrics ("Minimum Attribute Prediction Accuracy" vs "Minimum Fairness Gap": p=9.60e-94, one-tailed Wilcoxon rank-sum test; "Minimum Attribute Prediction AUROC" vs "Minimum Fairness Gap": p=1.95e-12, one-tailed Wilcoxon rank-sum test). **b,** We select the model for each algorithm with the minimum in-distribution fairness gap. We evaluate its OOD fairness against the oracle on the same 42 settings. We find that removing demographic encoding (i.e., DANN) leads to the best OOD fairness ("DANN" vs "ERM": p=1.86e-117, one-tailed Wilcoxon rank-sum test). Error bars indicate 95% confidence intervals estimated using non-parametric bootstrap sampling (n=1,000) are shown.

## Discussion

We have demonstrated the interplays between the demographic encoding of attributes as "shortcuts" in medical imaging AI models, and how they change under distribution shifts. Importantly, we were able to validate our findings across global-scale datasets in radiology (Table 1), and across multiple medical imaging modalities (Extended Data Figs. 1, 2). The results show that algorithmic encoding of protected attributes leads to unfairness (Fig. 2), and mitigating shortcuts can reduce ID fairness gaps and maintain performance (Fig. 3). However, our results also show that there exists inherent tradeoff for clinically meaningful metrics beyond fairness (Fig. 4), and such fairness does not transfer under distribution shift (Fig. 5). We provide initial strategies to dissect and explain the model fairness under distribution shifts (Fig. 6). Our results further reveal actionable algorithm and model selection strategies for out-of-domain fairness (Fig. 7).

Our results have multiple implications. First, they offer a cautionary tale on the efficacy and consequences of eliminating demographic shortcuts in disease classification models. On the one hand, removing shortcuts addresses ID fairness which is a crucial consideration in fair clinical decision making[14]. On the other hand, the resulting trade-offs with other metrics and non-transferability to OOD settings raises the question about the long-term utility in removing such shortcuts. This is particularly complex in the healthcare setting, where the relationship between the demographics and the disease or outcome label are complex[48], variables can be mislabeled[49,50], and distribution shifts between domains are difficult to quantify[1], much less predict.

Second, we frame demographic features as potential "shortcuts", which should not be utilized by the model to make disease predictions. However, some demographic variables could be a direct *causal* factor in some diseases (e.g., sex as a causal factor of breast cancer). In these cases, it would not be desirable to remove all demographic reliance, but instead match the reliance of the model on the demographic attribute to its true causal effect[51,52]. In the tasks we have examined here, demographic variables such as race may have an indirect effect on disease (e.g., through socioeconomic status)[53], which may vary across geographic location, or even time period[54]. Whether demographic variables should serve as proxies for these causal factors is a decision that should rest with the model deployers[16,48,55-57].

Third, we present a preliminary decomposition for diagnosing OOD model fairness changes, by expressing it as a function of the ID fairness gap, and the performance change of each group. We find that the disparate impacts of distribution shift on per-group performance is a significant contributor to lack of fairness in OOD settings. Our work suggests that, for practitioners trying to achieve fairness in models deployed in a different domain, mitigating ID fairness (e.g., through methods we have evaluated here) is at least as important as mitigating the impact of distribution shift for particular groups. However, building models robust to arbitrary domain shifts is, in general, a challenging task[58,59]. Having some knowledge or data about *how* the distributions may shift, or even the ability to actively collect data for particular groups, may be necessary[60,61]. Developing methods and deriving theoretical characterizations of fairness under distribution shift is an active area of research[46,47,62].

Fourth, the Food and Drug Administration (FDA) is currently the primary regulatory agency for new medical technologies[63]. The efficacy and safety of such models may also be regulated under the recent White House Executive Order on the safe, secure, and trustworthy development and use of artificial intelligence[64]. Under current processes, the efficacy of AI models in clinical settings is assessed by the product creator, and does not require external validation, e.g., by the FDA itself, or another body with access to out-of-distribution data from another source. Importantly, our results demonstrate that even if a model is locally optimal in a single source, it must be regularly evaluated for performance under shift[65-67]. While many experts have called for specific documentation of fairness evaluations that are performed for approved models[68], our finding that fairness does not transfer to OOD distributions challenges this popular opinion of a single fair model across different settings. Specifically, assurances by a developer that a model was fair at the time of testing is not a guarantee that this fairness will be maintained in new settings. This has implications for how to regulate fairness, a common area of interest for several federal agencies including The Centers for Medicare & Medicaid Services and The Joint Commission. Transparency on how the model performs across various subgroups using both fairness and other performance metrics may still be useful to implementers to adapt the model for their local implementations. Regardless, real world model surveillance of AI performance including fairness degradation in a specific setting may be more valuable than out of the box guarantees[69]. Finally, when a model is deployed in any clinical environment, both its overall and per-group performance, as well as associated clinical outcomes, should be continuously monitored[70].

Finally, while we imply that smaller "fairness gaps" are better, enforcing these group fairness definitions can lead to worse utility and performance for all groups[42,71-76], and other fairness definitions may be better suited to the clinical setting[9,77]. We encourage practitioners to choose a fairness definition that is best-suited to their use case, and carefully consider the performance-equality trade-off. In particular, we study bias through the lens of algorithmic fairness. The impact of minimizing algorithmic bias on real-world health disparities, the ultimate objective, is complex[78], and there is no guarantee that deploying a fair model will lead to equitable outcomes. In addition, though we construct several models for clinical risk prediction in this paper, we do not advocate for deployment of these models in real-world clinical settings without practitioners carefully testing models on their data and taking other considerations into account (e.g., privacy, regulation, interpretability)[1,3].

# Methods

**Datasets and Pre-Processing**

The datasets used in this study are summarized in Extended Data Table 1. Unless otherwise stated, we train models on MIMIC-CXR[23], and evaluate on an OOD dataset created by merging CheXpert[24], NIH[25], SIIM[26], PadChest[27], and VinDr[28]. We include all images (both frontal and lateral), and split each dataset into 70% train, 15% validation, 15% test sets. Note that only MIMIC-CXR and CheXpert have patient race information available. For MIMIC-CXR, demographic information was obtained by merging with MIMIC-IV[79]. For CheXpert, separate race labels were obtained from the Stanford AIMI website. Where applicable, we drop patients with missing values for any attribute.

For all datasets, we exclude samples where the corresponding patient has missing age or sex. For ODIR and ISIC, we drop samples from patients younger than 18 and older than 80 due to small sample sizes (i.e., smaller than 3% of the total dataset).

We scale all images to 224x224 for input to the model. We apply the following image augmentations during training only: random flipping of the images along the horizontal axis, random rotation of up to 10 degrees, and a crop of a random size (70% to 100%) and a random aspect ratio (3/4 to 4/3).

**Evaluation Methods**

To evaluate the performance of disease classification in medical imaging, we use the following metrics: the area under the ROC curve (AUC), True Positive Rate (TPR), True Negative Rate (TNR), and Expected Calibration Error (ECE).

The TPR and TNR are calculated as (TP: True Positive; FN: False Negative; TN: True Negative; FP: False Positive):

$$TPR = \frac{TP}{TP + FN}$$
$$TNR = \frac{TN}{TN + FP}$$

When reporting the sensitivity and specificity, we follow prior work[14,80] in selecting the threshold that maximizes the F1 score. This threshold optimization procedure is conducted separately for each dataset, task, algorithm, and attribute combination. We follow standard procedures to calculate the 95% confidence interval for sensitivity and specificity.

We also reported AUC, which is the area under the corresponding ROC curves showing an aggregate measure of detection performance. Finally, we report the Expected Calibration Error (ECE)[81], which we compute using the *netcal* library[82].

**Assessing the Fairness of ML Models**

In order to assess the fairness of ML models, we evaluate the metrics described above for each demographic group, as well as the difference in the value of the metric between groups. Equality of TPR and TNR between demographic groups is known in the algorithmic fairness literature as equal odds[83]. As the models we study in this work are likely to be used as screening or triage tools, the cost of a False Positive (FP) may be different from the cost of a False Negative (FN). In particular, for No Finding prediction, FPs (corresponding to underdiagnosis[14]) would be more costly than FNs, and so we focus on the FPR (or TNR) for this task. For all remaining disease prediction tasks, we focus on the FNR (or TPR) for the same reason. Equality in one of the class conditioned error rates is an instance of equal opportunity[32].

Finally, we also examine the per-group ECE and ECE gap between groups. Note that zero ECE for both groups (i.e., calibration per group) implies the fairness definition known as sufficiency of the risk score[83]. We emphasize that differences in calibration between groups is a significant source of disparity, as consistent under or over-estimation of risk for a particular group could lead to under or over-treatment for that group at a fixed operating threshold relative to the true risk[84].

**Training Details**

We train DenseNet-121[33] models on each task, initializing with ImageNet[85] pretrained weights. We evaluate six algorithms: empirical risk minimization (ERM[34]), resampling to equalize group size (Resample[35]), group distributionally robust optimization (GroupDRO[36]), domain adversarial training (DANN[37]), domain adversarial training conditioned on the label (CDANN[38]), and weight averaging (MA[39]).

For each combination of task, algorithm, and demographic attribute, we conduct a random hyperparameter search[86] with 15 runs. During training, for a particular attribute, we evaluate the validation set worst-group validation AUROC every 1,000 steps, and early stop if this metric has not improved for 5 evaluations. We tune the learning rate and weight decay for all algorithms, and also tune algorithm specific hyperparameters as mentioned in the original works. We select the hyperparameter setting that maximizes the worst-attribute validation AUROC. Confidence intervals are computed as the standard deviation across three different random seeds for each hyperparameter setting.

To obtain the level of demographic encoding within representations (Fig. 2), we first compute representations using a trained disease prediction model. We freeze these representations, and train a multi-class multinomial logistic regression model to predict the demographic group using the training set using the scikit-learn library[87]. We vary the L2 regularization strength between $10^{-5}$ and 10, and select the model with the best macro-averaged AUROC on the validation set. We report the macro-averaged AUROC on the test-set.

**Decomposing Out-of-Distribution Fairness**

Here, we present a first approach towards decomposing the fairness gap in an out-of-distribution environment as a function of the in-distribution fairness gap, and the impact that the distribution shift has in each group. In particular, let $D_{src}$ and $D_{tar}$ be the source and target datasets, respectively. Let $g \in G$ be a particular group from a set of groups. Let $L_f(g, D)$ be an evaluation metric for a model $f$, which is decomposable over individual samples, i.e., $L_f(g, D) = \sum_{(x,y,g') \in D; g' = g} l(f(x), y)$. Examples of such metrics are the accuracy, TPR, or TNR. Then, we can decompose:

$$L_f(g_1, D_{tar}) - L_f(g_2, D_{tar}) = [L_f(g_1, D_{src}) - L_f(g_2, D_{src})] + [L_f(g_2, D_{src}) - L_f(g_2, D_{tar})] - [L_f(g_1, D_{src}) - L_f(g_1, D_{tar})].$$

The left-hand term is the fairness gap in the out-distribution environment, and the three terms on the right are (1) the fairness gap in the in-distribution data, (2) the impact of the distribution shift on $g_2$, and (3) the impact of the distribution shift on $g_1$. We note that to achieve a low fairness gap in the out-of-distribution environment, it is important not only to minimize the in-distribution fairness gap (term 1), but also to minimize the difference in how the distribution shift impacts each group (term 2 - term 3).

**Evaluation with Different Medical Imaging Modalities**

In addition to radiology, we also examine medical AI applications in dermatology and ophthalmology to corroborate our findings. Specifically, Extended Data Fig. 1 shows the results for dermatological imaging. We use the ISIC dataset[30], which contains 32,259 images sourced from multiple international sites. We focus on the "No Finding" task, taking into account "sex" and "age" as the sensitive demographic attributes (Extended Data Fig. 1a). Similar to our observations in radiology, we identify fairness gaps within subgroups based on age and sex (Extended Data Fig. 1b), although these disparities are less significant than those observed in chest X-ray assessments (e.g., fairness gaps smaller than 2%). This was further confirmed by the Pareto front plot, where most models including ERM could achieve good performance-fairness tradeoff (Extended Data Fig. 1c).

We extend our analysis to ophthalmology images, specifically focusing on retinopathy detection, using the ODIR dataset[31] with 6,800 images (Extended Data Fig. 2). The task we consider is "Retinopathy", with "sex" and "age" being used as demographic attributes (Extended Data Fig. 2a). Notably, significant subgroup fairness gaps are observed in age (43% FNR gap between groups "60-80" and "18-40"). In contrast, the fairness gap based on sex is less significant, with a 3% FNR difference between "female" and "male" subgroups. We further verify the Pareto front for both attributes, where similar observations hold that algorithms for fixing demographic shortcuts could improve in-distribution fairness while incurring minimal detriments to performance (measure in AUROC).

**Analysis on Underdiagnosis versus Overdiagnosis**

In evaluating fairness metrics, our primary study centered on underdiagnosis, specifically the disparities in FPR for "No Finding" and discrepancies in FNR for other conditions. However, an alternative approach involves focusing on overdiagnosis, defined as variances in FNR for "No Finding" and differences in FPR for other diseases. We present findings between their relationship in Extended Data Fig. 3. An analysis spanning two datasets (MIMIC and CheXpert) and various tasks revealed a consistent pattern: larger gaps in underdiagnosis tend to correspond with more significant overdiagnosis discrepancies. Nonetheless, certain task and attribute combinations exhibit more complex trends, indicating a necessity for deeper exploration and informed decision-making regarding the most appropriate fairness metrics for critical disease evaluations in practical medical settings.

**Statistical Analysis**

**Correlation.** To calculate the correlations between variables, we used Pearson correlation coefficients and their associated p-value (two-sided t-test, $\alpha=0.05$). 95% CI for the Pearson correlation coefficient was calculated.

**Increase in OOD fairness gap.** One-tailed Wilcoxon rank-sum test ($\alpha=0.05$) was used to assess the increase in OOD fairness gap compared to oracle models.

**Confidence intervals.** We use the non-parametric bootstrap sampling to generate confidence intervals: random samples of size n (equal to the size of the original dataset) are repeatedly sampled 1,000 times from the original dataset with replacement. We then estimate the increase in OOD fairness gap compared to oracle using each bootstrap sample ($\alpha=0.05$).

All statistical analysis was performed with Python version 3.9 (Python Software Foundation).

**Data Availability**

All datasets used in this study are publicly available. The MIMIC-CXR[23] and VinDr-CXR[28] datasets are available from PhysioNet after the completion of a data use agreement and a credentialing procedure. The CheXpert[24] dataset, along with associated race labels, is available from the Stanford AIMI website. The ChestX-ray14[25] (NIH) dataset is available to download from the National Institute of Health Clinical Center. The PadChest[27] dataset can be downloaded from the Medical Imaging Databank of the Valencia Region. The SIIM-ACR[26] Pneumothorax Segmentation dataset can be downloaded from its Kaggle contest page. The ISIC[30] 2020 dataset can be downloaded from the SIIM-ISIC Melanoma Classification Challenge page. The ODIR[31] dataset can be obtained from the ODIR 2019 challenge hosted by Grand Challenges.

**Code Availability**

Code that supports the findings of this study is publicly available with an open-source license at https://github.com/YyzHarry/shortcut-ood-fairness.

**Author Contributions**

Y.Y., H.Z., and M.G. conceived and designed the study. Y.Y. and H.Z. performed data collection, processing, and experimental analysis. Y.Y., H.Z., J.W.G., D.K., and M.G. interpreted experimental results and provided feedback on the study. Y.Y., H.Z., J.W.G., and M.G. wrote the original manuscript. M.G. supervised the research. All authors reviewed and approved the manuscript.

**Competing Interests**

D.K. is a cofounder of Emerald Innovations, Inc., and serves on the scientific advisory board of Janssen and the data and analytics advisory board of Amgen. The remaining authors declare no competing interests.

# Extended Data

**Extended Data Table 1. Prevalence rates for each demographic subgroup of the datasets used for training models in this study.**

|  |  | MIMIC | | | | CheXpert | | | |
|---|---|---|---|---|---|---|---|---|---|
|  |  | Cardiomegaly | No Finding | Effusion | Pneumothorax | Cardiomegaly | No Finding | Effusion | Pneumothorax |
| Sex (%) | Female | 15.1 | 42.6 | 18.9 | 2.8 | 11.6 | 10.2 | 38.8 | 8.3 |
|  | Male | 14.8 | 37.2 | 21.1 | 4.0 | 12.4 | 9.9 | 38.4 | 9.0 |
| Race (%) | Asian | 16.6 | 36.0 | 24.2 | 5.4 | 12.7 | 10.4 | 40.5 | 9.8 |
|  | Black | 17.6 | 44.3 | 13.4 | 1.8 | 19.6 | 11.7 | 31.7 | 5.8 |
|  | White | 15.5 | 34.6 | 24.0 | 4.0 | 11.5 | 9.4 | 39.4 | 9.1 |
|  | Other | 11.1 | 52.5 | 12.6 | 2.5 | 11.7 | 10.8 | 37.6 | 8.0 |
| Age (%) | 18-40 | 6.8 | 64.0 | 8.1 | 3.6 | 9.1 | 20.5 | 27.0 | 12.5 |
|  | 40-60 | 11.4 | 46.5 | 15.0 | 3.0 | 10.1 | 12.4 | 36.2 | 8.6 |
|  | 60-80 | 17.6 | 32.5 | 23.9 | 3.8 | 12.4 | 7.0 | 42.3 | 8.9 |
|  | 80-100 | 22.9 | 23.3 | 31.0 | 3.0 | 17.9 | 3.7 | 44.2 | 5.0 |
| Intersection (%) | Asian Female | 16.9 | 38.6 | 22.7 | 4.7 | 12.5 | 10.7 | 40.5 | 9.8 |
|  | Asian Male | 16.4 | 33.6 | 25.5 | 6.1 | 12.8 | 10.1 | 40.4 | 9.8 |
|  | Black Female | 18.3 | 46.6 | 13.0 | 1.5 | 20.9 | 11.5 | 31.8 | 4.7 |
|  | Black Male | 16.7 | 41.0 | 13.9 | 2.2 | 18.2 | 11.9 | 31.6 | 6.8 |
|  | White Female | 15.5 | 36.3 | 23.5 | 3.5 | 10.2 | 9.4 | 39.9 | 9.0 |
|  | White Male | 15.4 | 33.3 | 24.4 | 4.4 | 12.4 | 9.4 | 39.0 | 9.2 |
|  | Others Female | 10.7 | 57.2 | 10.9 | 1.7 | 12.1 | 11.4 | 37.6 | 7.0 |
|  | Others Male | 11.5 | 48.1 | 14.2 | 3.3 | 11.4 | 10.4 | 37.6 | 8.7 |

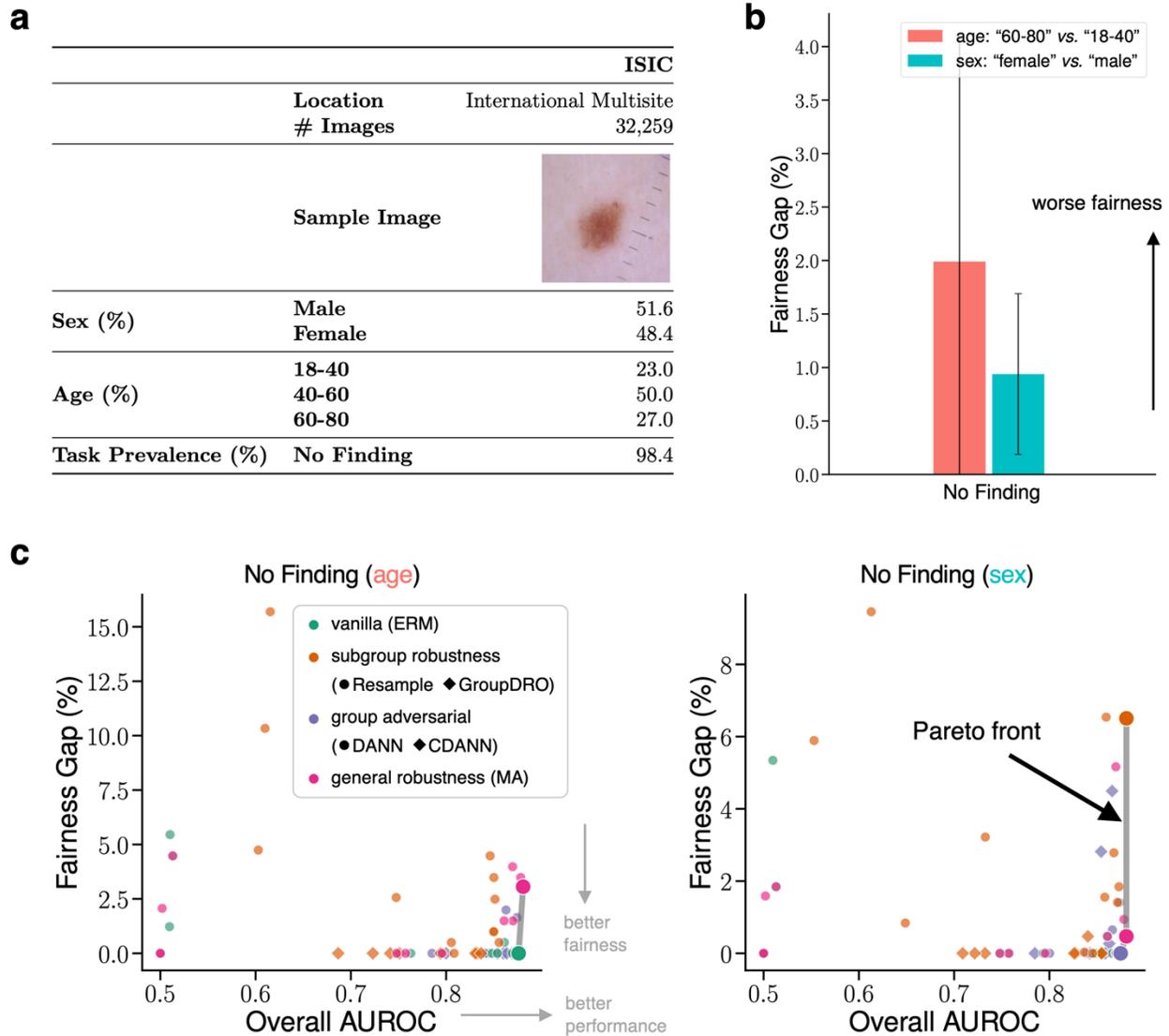

**Extended Data Figure 1. Evaluation results for models on the ISIC dataset for No Finding prediction. a,** Dataset statistics. **b,** Subgroup fairness gaps of the ERM model as defined by FPR. Each subgroup contains at least 100 samples for analysis (age: subgroup "60-80" vs "18-40"; sex: subgroup "female" vs "male"). We find that disparities in FPR are small and statistically insignificant in the case of age. **c,** Trade-off between the fairness gap and overall AUROC for all trained models, evaluated against sensitive attribute age and sex, respectively. We find that most models, including ERM, achieve a good fairness-performance tradeoff.

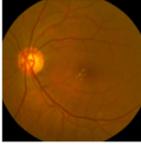
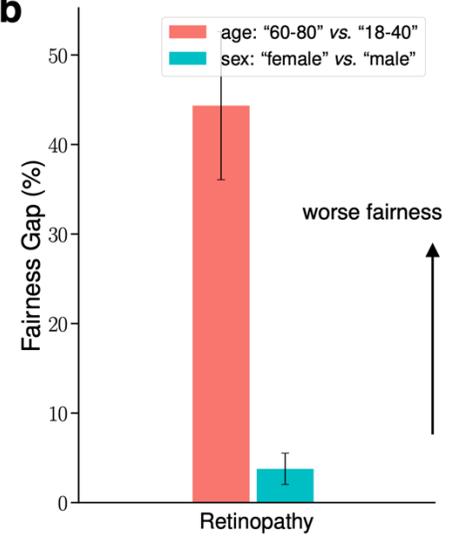
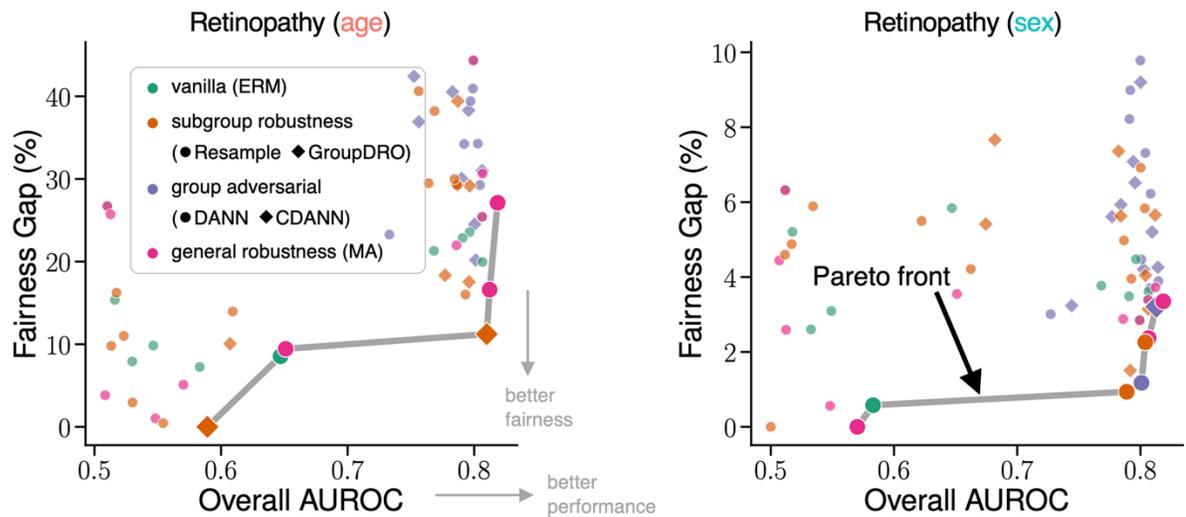

**Extended Data Figure 2. Evaluation results for models on the ODIR dataset for retinopathy prediction. a,** Dataset statistics. **b,** Subgroup fairness gaps of the ERM model as defined by FNR. Each subgroup contains at least 100 samples for analysis (age: subgroup "60-80" vs "18-40"; sex: subgroup "female" vs "male"). We find a significant FNR gap between age groups. **c,** Trade-off between the fairness gap and overall AUROC for all trained models, evaluated against sensitive attribute age and sex, respectively. We find, similar to the chest X-ray setting, that algorithms for fixing demographic shortcuts could improve in-distribution fairness while incurring minimal detriments to performance.

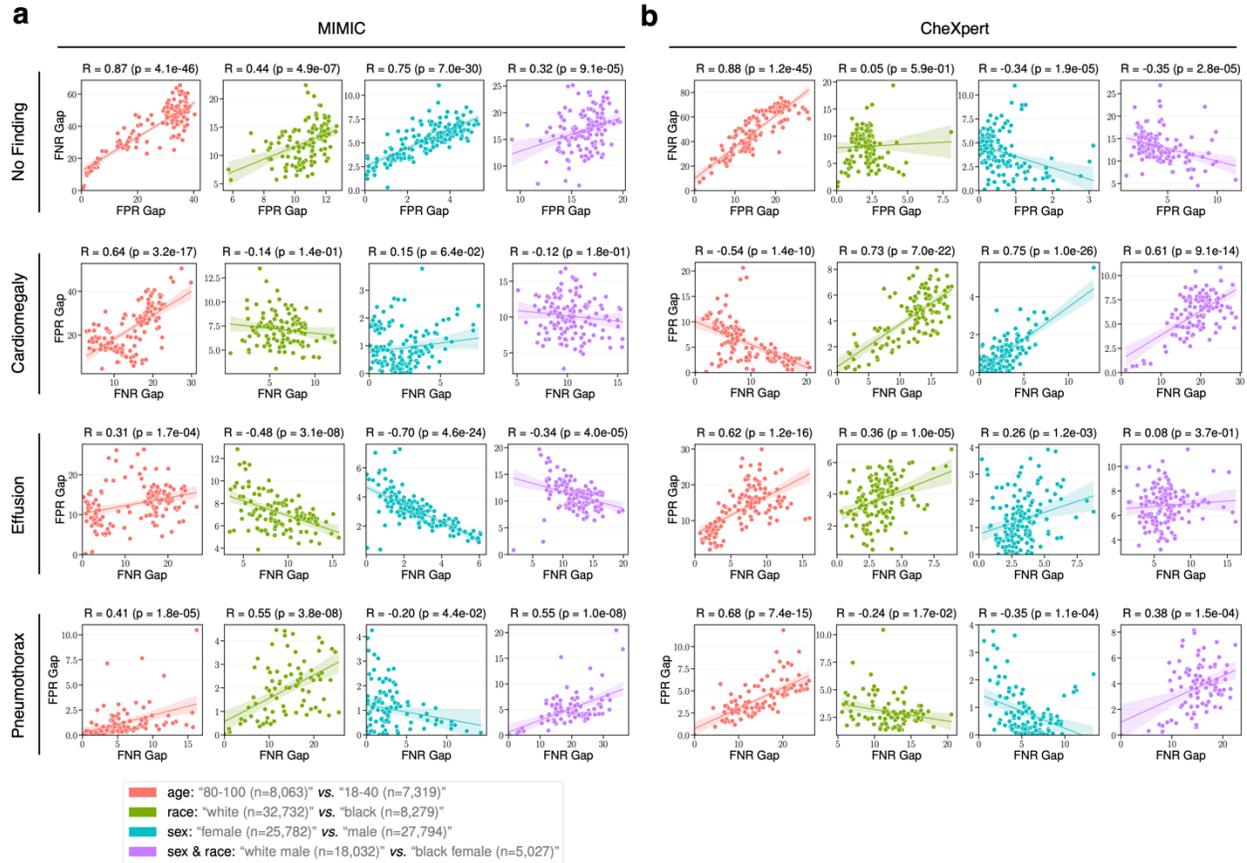

**Extended Data Figure 3. Trade-offs between the FPR gap and the FNR gap for each task and attribute, for models trained on MIMIC-CXR or CheXpert and evaluated on the same dataset for (a) No Finding and (b) Effusion prediction.** We evaluate these metrics across age ("80-100" vs "18-40"), sex ("female" vs male"), race ("White" vs. "Black"), and the intersection of sex and race ("White male" vs. "Black female"). We find for the most part, there is a positive correlation, indicating that fairer models achieve fairness with respect to both FPR and FNR (i.e., equal odds).

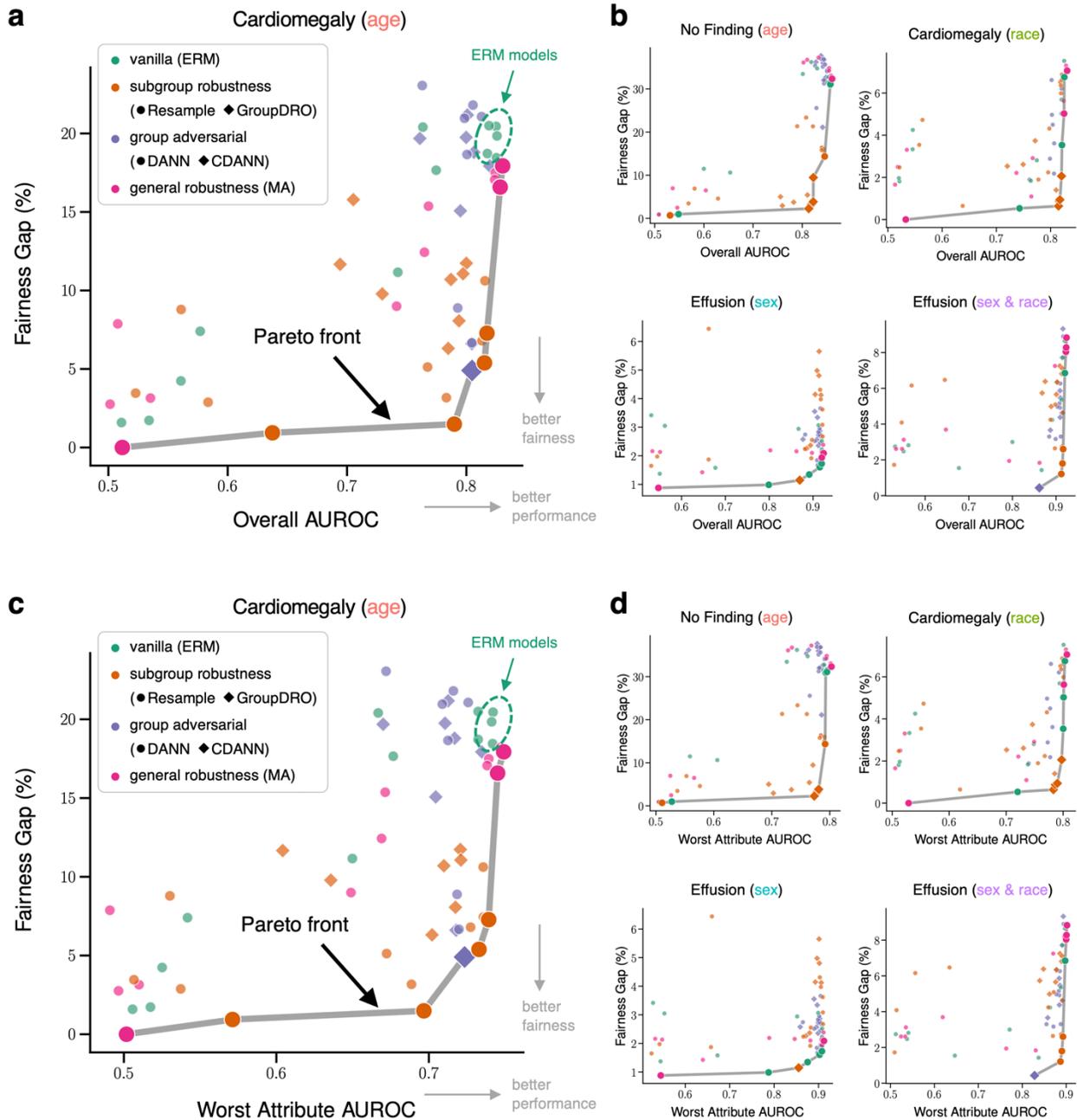

**Extended Data Figure 4. Algorithms for removing demographic shortcuts mitigate in-distribution fairness gaps and maintain performance. a, b,** Trade-off between the fairness gap and overall AUROC for all trained models. **c, d,** Trade-off between the fairness gap and the worst-group AUROC for all trained models. Each plot represents a specific disease prediction task (e.g., Cardiomegaly) with a specific attribute (e.g., age). In each case, we plot the Pareto front, the best achievable fairness gap with a minimum constraint on the performance.

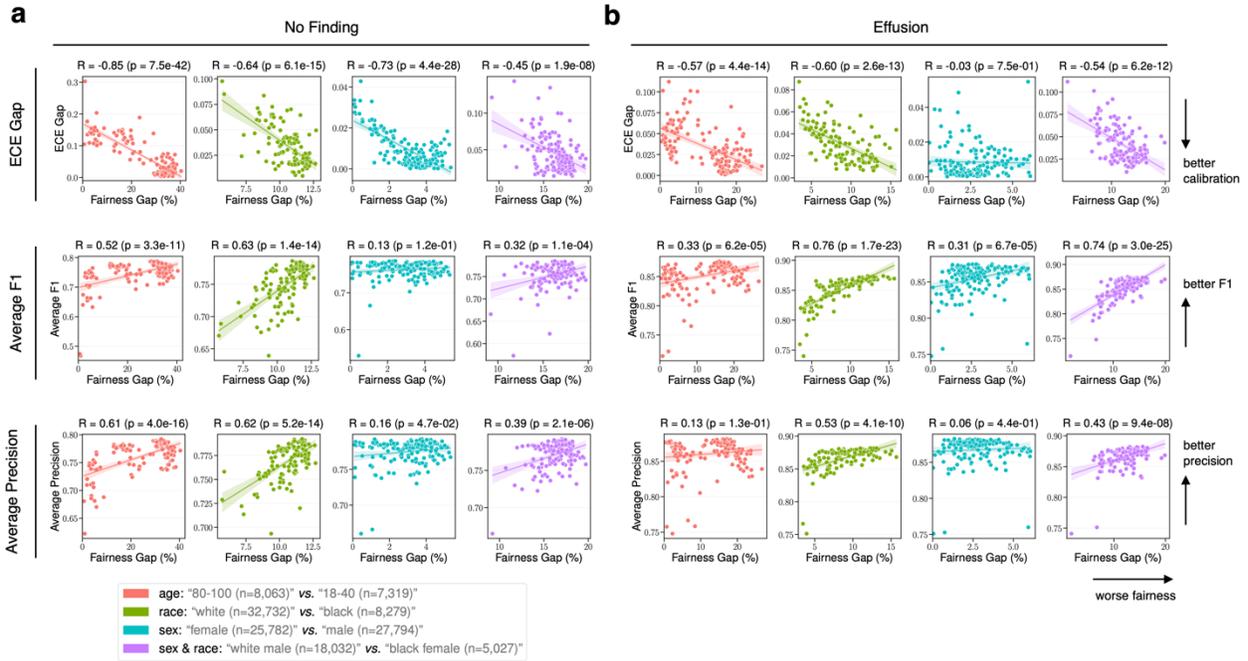

**Extended Data Figure 5. Trade-offs between the fairness gap and the ECE gap, Average F1, and Average precision, for models trained and evaluated on MIMIC-CXR for (a) No Finding and (b) Effusion prediction.** We evaluate these metrics across age ("80-100" vs "18-40"), sex ("female" vs male"), race ("White" vs. "Black"), and the intersection of sex and race ("White male" vs. "Black female"). We find that fairer models tend to exhibit larger ECE gaps, worse Average F1, and worse Average precision, indicating the undesirable tradeoff between fairness with other performance and calibration metrics.

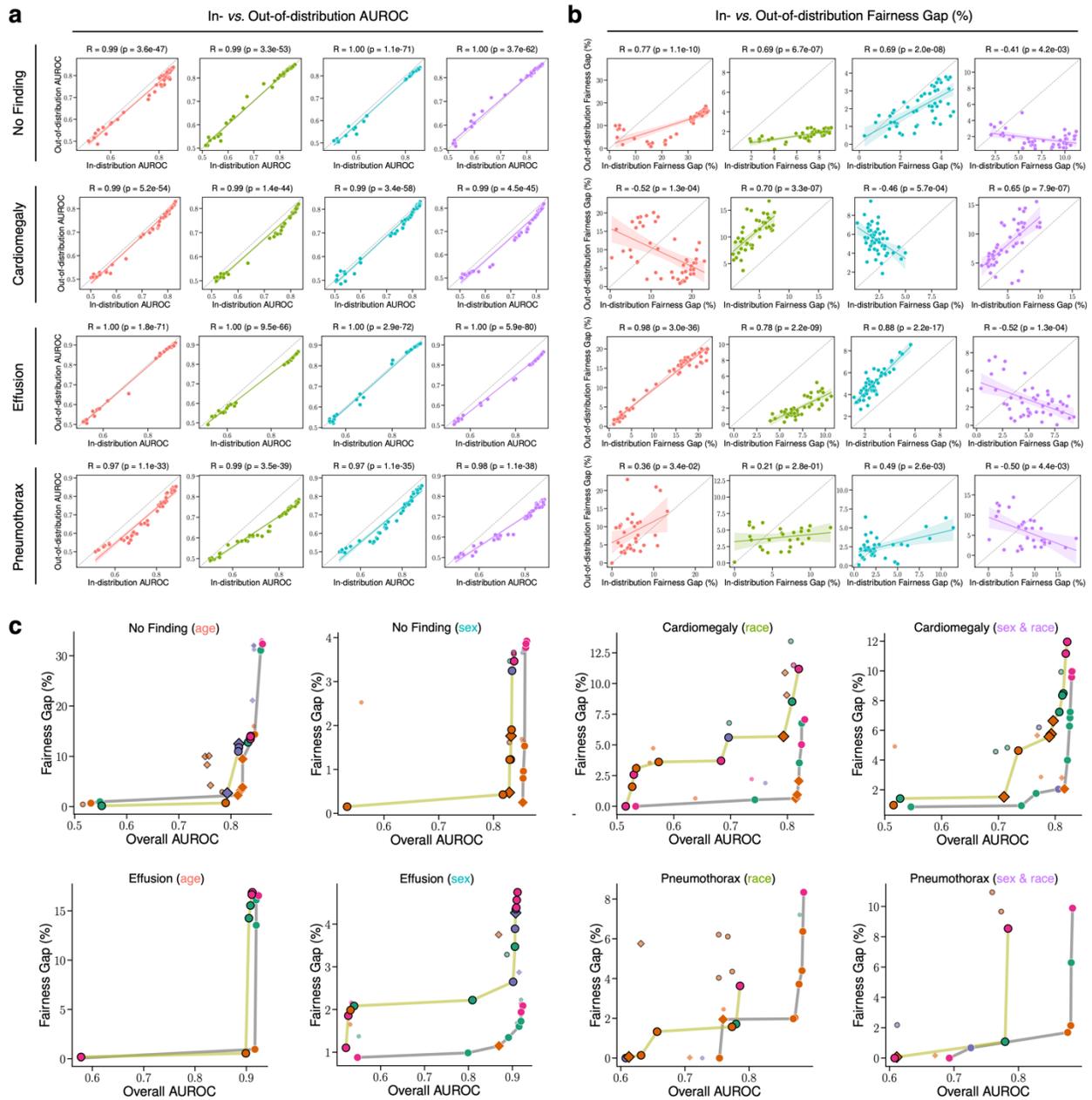

**Extended Data Figure 6. The transfer of performance (overall AUROC) and fairness between the ID (MIMIC-CXR) and OOD datasets, supplementing Figure 5. a**, We plot the OOD performance versus the ID performance for each task and attribute combination, **b,** We plot the OOD fairness versus the ID fairness for each task and attribute combination. **c,** We plot the Pareto front between fairness and performance for ID and OOD, finding that models that are Pareto optimal ID often do not maintain Pareto optimality OOD.

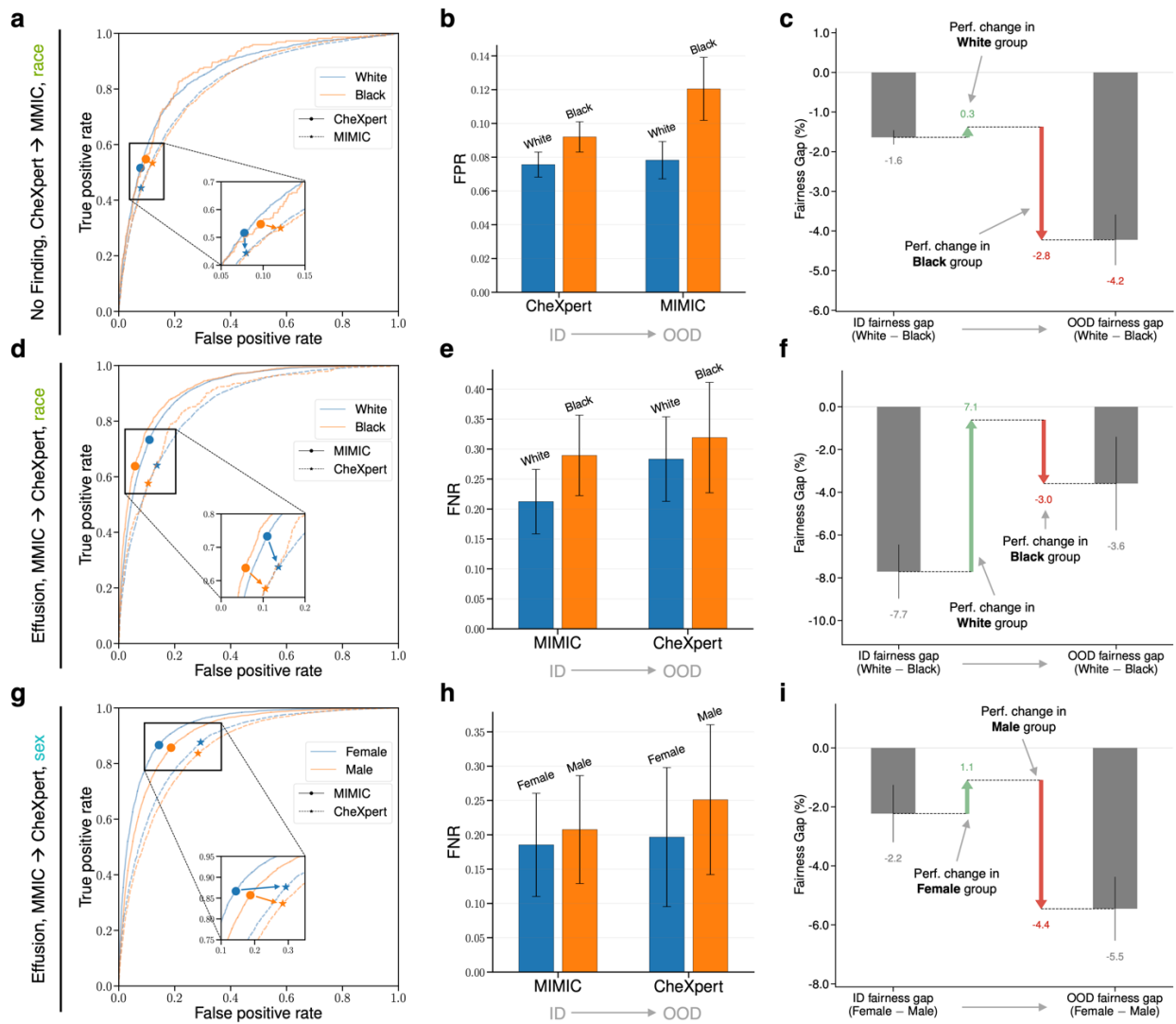

**Extended Data Figure 7. Examining the behavior of specific ERM models, supplementing Figure 6. a, d, g,** We plot the receiver operating characteristic (ROC) curves for each group and each dataset, marking the operating point of the model. **b, e, h,** This shift in the operating point results in a change in the FPR and FNR values of groups on the OOD dataset. **c, f, i,** We decompose the OOD fairness gap as a function of the ID fairness gap, and the change in FPR for each of the groups.